# A HYBRID INTELLIGENT MODEL FOR SOFTWARE COST ESTIMATION

**Wei Lin Du[1], Luiz Fernando Capretz[2], Ali Bou Nassif[2], Danny Ho[3]**

[1]DRN e-Commerce, London, Ontario, Canada;
[2]Department of ECE, University of Western Ontario, London, Ontario, Canada
[3]NFA Estimation Inc., Richmond Hill, Ontario, Canada
Email: lcapretz@uwo.ca

## ABSTRACT

Accurate software development effort estimation is critical to the success of software projects. Although many techniques and algorithmic models have been developed and implemented by practitioners, accurate software development effort prediction is still a challenging endeavor in the field of software engineering, especially in handling uncertain and imprecise inputs and collinear characteristics. In this paper, a hybrid intelligent model combining a neural network model integrated with fuzzy model (neuro-fuzzy model) has been used to improve the accuracy of estimating software cost. The performance of the proposed model is assessed by designing and conducting evaluation with published project and industrial data. Results have shown that the proposed model demonstrates the ability of improving the estimation accuracy by 18% based on the Mean Magnitude of Relative Error (MMRE) criterion.

## 1. INTRODUCTION

On-time delivery, budget control and high quality products are critical goals for software project management. The cost, quality and delivery of software projects are affected by the accuracy of software effort estimation (Nassif et al., 2010). Software engineering practices have specific characteristics that differentiate this field from traditional engineering. In particular, various factors affect software effort estimation in organizations and projects, including inconsistent software processes and measurement definitions in projects, substantial diversity among projects, and extreme differences in product sizes. Consequently, these situations create challenges in the practice of software effort estimation, making it difficult to yield a high degree of accuracy in estimation. Many studies have focused on developing software cost estimation models and techniques. These include algorithmic models, such as COCOMO (Boehm, 1981) (Briand and Wieczorek, 2002), SLIM (Putnam, 1978), SEER-SEM (Galorath and Evans, 2006), machine learning techniques. These models and techniques have been introduced and used in the software industry. However, modeling accuracy affects the quality of estimation. Hence, these studies are aimed at improving the predictive performance of current models by introducing new techniques and methodologies.

SEER-SEM (Galorath and Evans, 2006) appeals to software practitioners because of its powerful estimation features. It has been developed with a combination of estimation functions for performing various estimations. Created specifically for software effort estimation, the SEER-SEM model was influenced by the framework of Putnam (Putnam, 1978). As one of the algorithmic estimation models, SEER-SEM has two main limitations on effort estimation. First, there are over fifty input parameters related to the various factors of software projects, which might increase the complexity of SEER-SEM, especially for managing the uncertainty from these inputs. Second, the specific details of SEER-SEM increase the difficulty of discovering the non-linear relationship between the parameter inputs and the corresponding outputs. Overall, these two major limitations can lead to a lower accuracy in effort estimation by SEER-SEM. This research attempts to resolve the main limitation of the SEER-SEM effort estimation model. For accurately estimating software effort, neural network and fuzzy logic approaches are adopted to create a neuro-fuzzy model, which is subsequently combined with SEER-SEM. The Adaptive Neuro-Fuzzy Inference System (ANFIS) (Jang, 1993) is used as the architecture of each neuro-fuzzy sub-model.

Some researchers have used machine learning techniques to improve the accuracy of software cost estimation. This includes (Huang et al., 2007) and (Huang et al., 2004) who used a neuro-fuzzy model to improve the accuracy of the COCOMO Model, other work such as (Nassif et al., 2013), (Nassif et al., 2012) and (Nassif et al., 2011) have been used to improve the accuracy of the Use Case Point Model using Machine Learning techniques and (Du et al., 2010) who used a neural network with fuzzy logic model to improve the SEER-SEM algorithm; however, the evaluation conducted in the latter work was poor.

In this work, the proposed model is evaluated using a cross-validation technique on published industrial data. Experiments have shown that our model surpasses the SEER-SEM model by 18% based on the Mean Magnitude of Relative Error (MMRE) criterion. Our model also outperforms the SEER-SEM model using other evaluation criteria such as MdMRE, PRED(0.3), PRED(0.5) and MSE but the most significant improvement was based on the MMRE criterion.

The remainder of the paper is organized as follows: Section 2 describes the proposed hybrid intelligent model. The evaluation of the model is presented in Section 3. Section 4 highlights the threats that might have deteriorated the validity of our model. Finally, Section 5 concludes the paper.

## 2. A HYBRID INTELLIGENT MODEL FOR SEER-SEM

### 2.1. SEER-SEM Model

The SEER-SEM model was proposed by Galorath in 1988 (Galorath and Evans, 2006). This model was motivated by the Putnam's model (SLIM) and the COCOMO model. The main inputs and outputs of the SEER-SEM model are depicted in **Fig. 1.**

The SEER-SEM effort estimation is calculated by the following equation:

$$E = 0.393469 \times K.$$

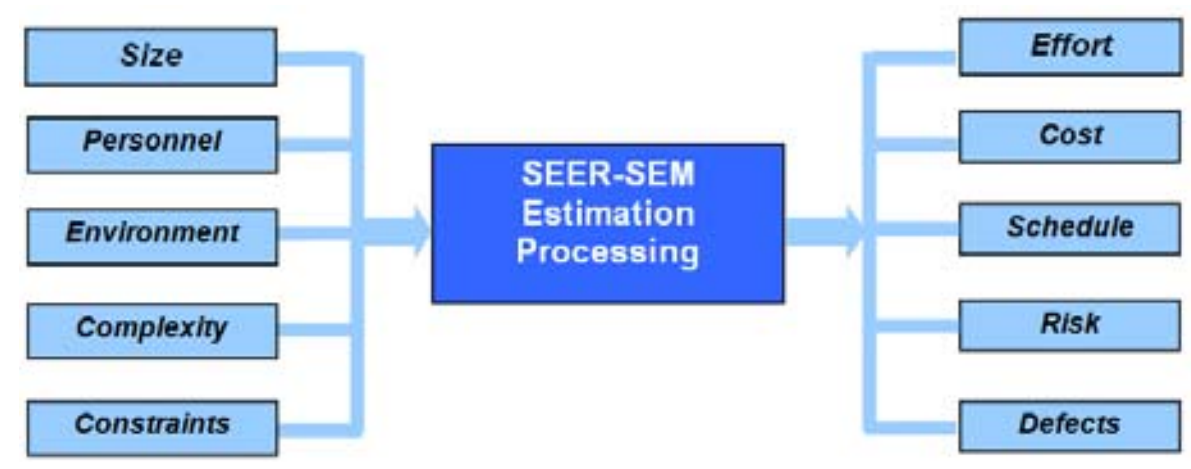

**Fig. 1.** Inputs and Outputs of the SEER-SEM Model

Where $E$ is the development effort in persons-year and $K$ is the total Life-cycle effort including development and maintenance (in person-years). K is directly proportional to staffing complexity and software size (KLOC) and inversely proportional to the effective technology used to develop the project.

### 2.2. Neuro-Fuzzy Model

The structure of the hybrid model used in this paper is composed of inputs related to SEER-SEM algorithm, a neuro-fuzzy bank, corresponding values of inputs, an algorithmic model (SEER-SEM in this case, but any algorithmic model can fit here), and outputs for effort estimation. The algorithmic model with the neuro-fuzzy bank can be considered as the major parts of the proposed model. The inputs of the proposed model are rating levels, which can be linguistic terms such as Low, Nominal, or High or continuous values. The main structure of the proposed model is depicted in **Fig. 2.**

Where $PR_i$ are the inputs of the SEER-SEM model and $NF_i$ are the neuro-fuzzy sub-models as shown in **Fig. 3.**

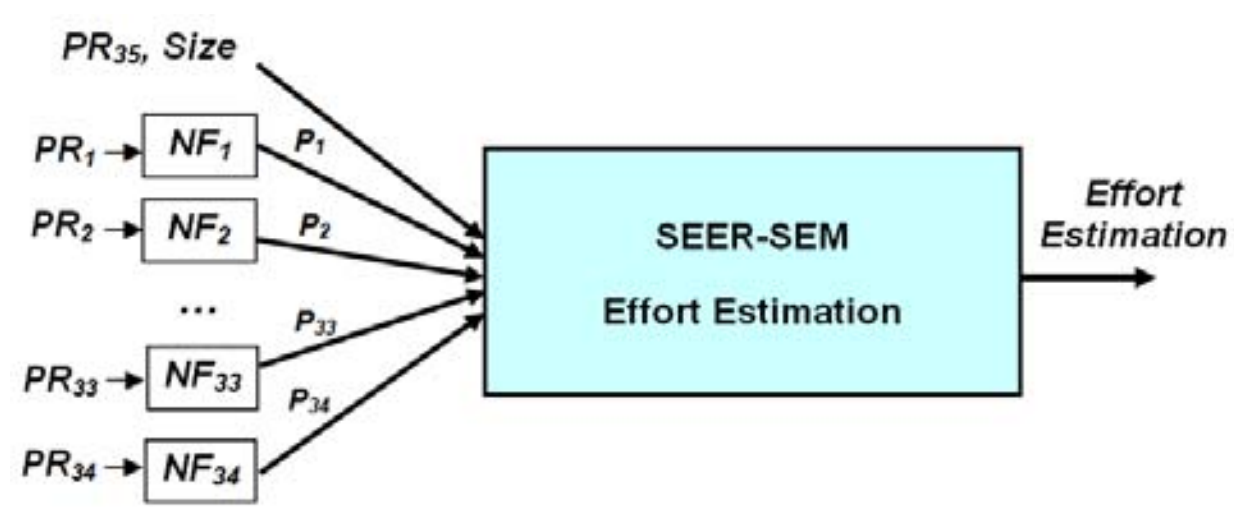

**Fig. 2.** Neuro-Fuzzy Model with SEER-SEM

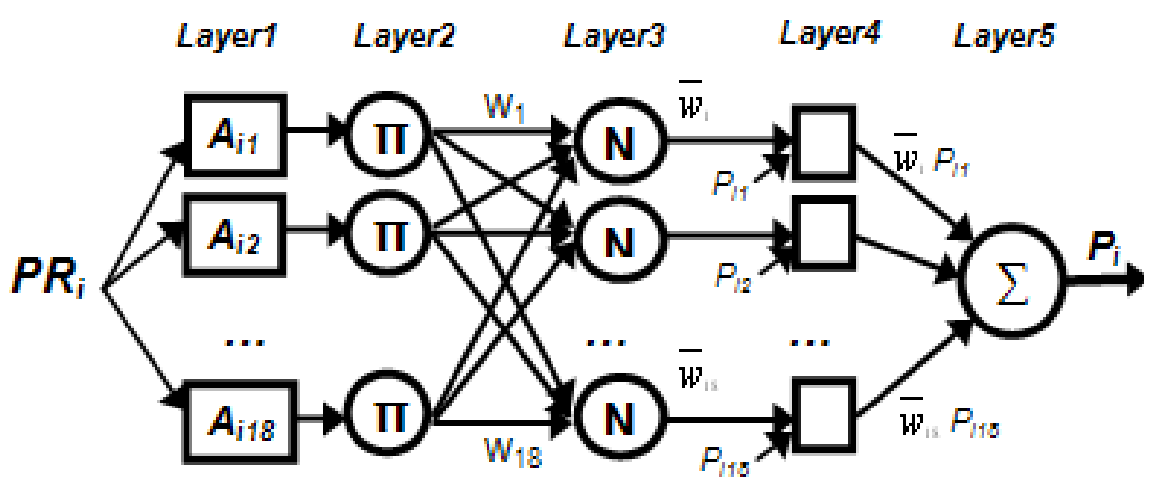

**Fig.3.** NFi model

## 3. MODEL EVALUATION

After incorporating the neuro-fuzzy model with SEER-SEM in the previous section, this section evaluates the proposed model by using industrial project data points. In our research, 99 project data points are used to train and test the performance of the proposed model. Among them, 93 published NASA project data points are from 6 centers and categorized to three development modes: embedded, organic, and semidetached. The rest are 6 industrial project data points (Panlilio-Yap and Ho, 1994). COCOMO 81 projects were transformed to COCOMO II then to SEER-SEM. The matching between SEER-SEM parameters and COCOMO drivers is depicted in **Appendix A.**

To assess the accuracy of the proposed model, we have used common evaluation criteria used in software estimation which are MMRE, MdMRE, PRED(x) and MSE.

- *MMRE:* This is a very common criterion used to evaluate software cost estimation models (Briand et al., 1999). The Magnitude of Relative Error (MRE) for each observation i can be obtained as:

$$MRE_i = \frac{|Actual\ Effort_i - Predicted\ Effort_i\ |}{Actual\ Effort_i}$$

MMRE can be achieved through averaging the summation of MRE over $N$ observations:

$$MMRE = \frac{1}{N}\sum_{i=1}^{N} MRE_i$$

MMRE is a common method used for evaluation prediction models; however, this method has been criticized by others such as (Foss et al., 2003), (Shepperd and Schofield, 1997) and (Myrtveit and Stensrud, 2012). For this reason, we used a statistical significant test to compare between the median of two samples based on the residuals. Since the residuals were not normally distributed, the non-parametric statistical test Mann-Whitney U has been used to assess the statistical significance between different prediction models.

- *MdMRE*: One of the disadvantages of the MRE is that it is sensitive to outliers. MdMRE has been used as another criterion because it is less sensitive to outliers.

$$MdMRE = median(MRE_i)$$

- *PRED(x)*: The prediction level (PRED) is used as a complimentary criterion to MMRE. PRED calculates the ratio of a project's MMRE that falls into the selected range (x) out of the total projects.

PRED (x) can be described as:

$$PRED(x) = \frac{k}{n}.$$

where k is the number of projects where $MRE_i \leq x$ and $n$ is the total number of observations. In this work, PRED(0.30) and PRED(0.50) have been used.

- *MSE*: The Mean Squared Error (MSE) is the mean of the square of the differences between the actual and the predicted efforts.

$$MSE = \frac{\sum_{i=1}^{N}(Actual_Effort_i - Estimated_Effort_i)^2}{N}.$$

The estimation accuracy is directly proportional to PRED (x) and inversely proportional to MMRE, MdMRE and MSE.

Experiments were conducted using the cross-validation technique to compare the original SEER-SEM model with the proposed neuro-fuzzy model (**Fig. 2**). The inputs of the models are software size and a set of parameters as explained in Section 2. The output of the models is software effort. The results of the evaluation criteria (MMRE, MdMRE, PRED and MSE), as well as the Mann-Whitney U test are reported in **Table 1.** The interval plot at 95% confidence level of the MMRE and the Boxplot are shown in **Fig. 4** and **Fig. 5**, respectively.

**Table 1.** Results of the Model Evaluation

| | MMRE | MdMRE | PRED(0.30) | PRED(0.50) | MSE | Mann-Whitney U (p value) |
|---|---|---|---|---|---|---|
| SEER-SEM | 0.57 | 0.27 | 52 | 66.25 | 287180 | 0.0183 |
| Neuro-fuzzy SEER-SEM | 0.39 | 0.24 | 55 | 71.25 | 261332 | |
| Improvement | 18% | 3% | 3% | 5% | 25848 | |

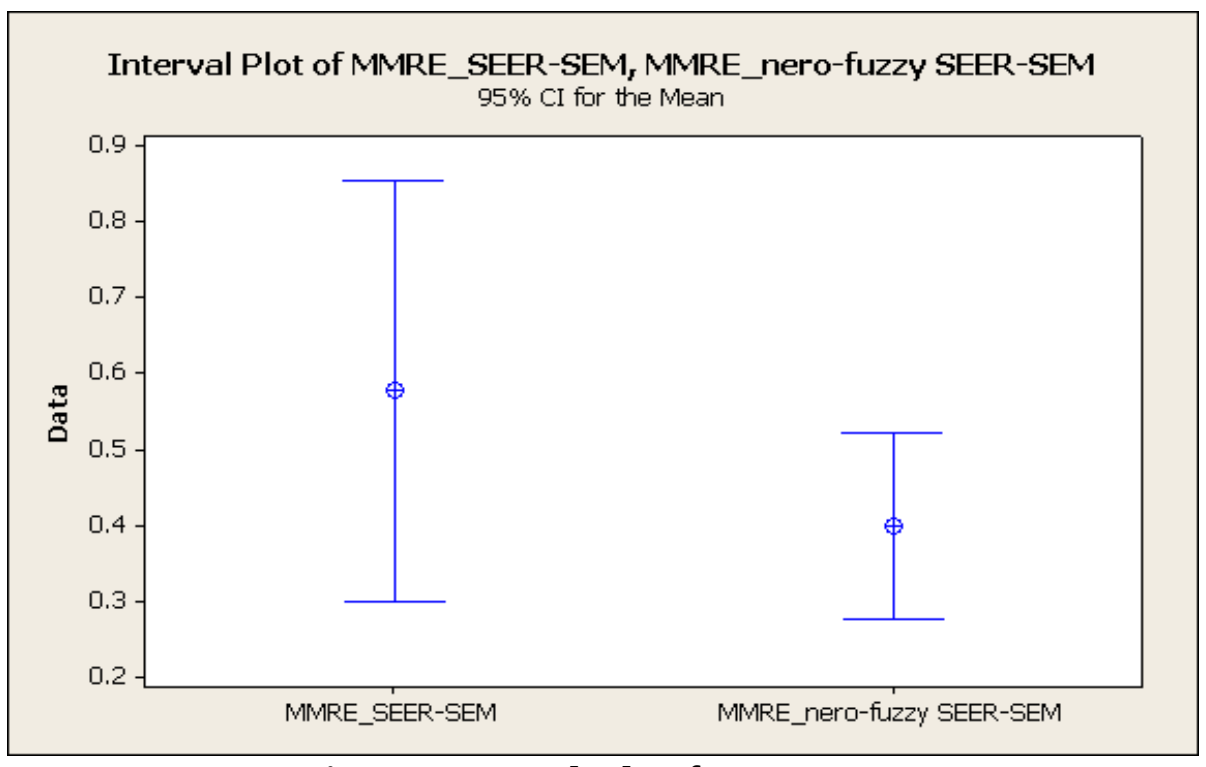

**Fig. 4.** Interval Plot for MMRE

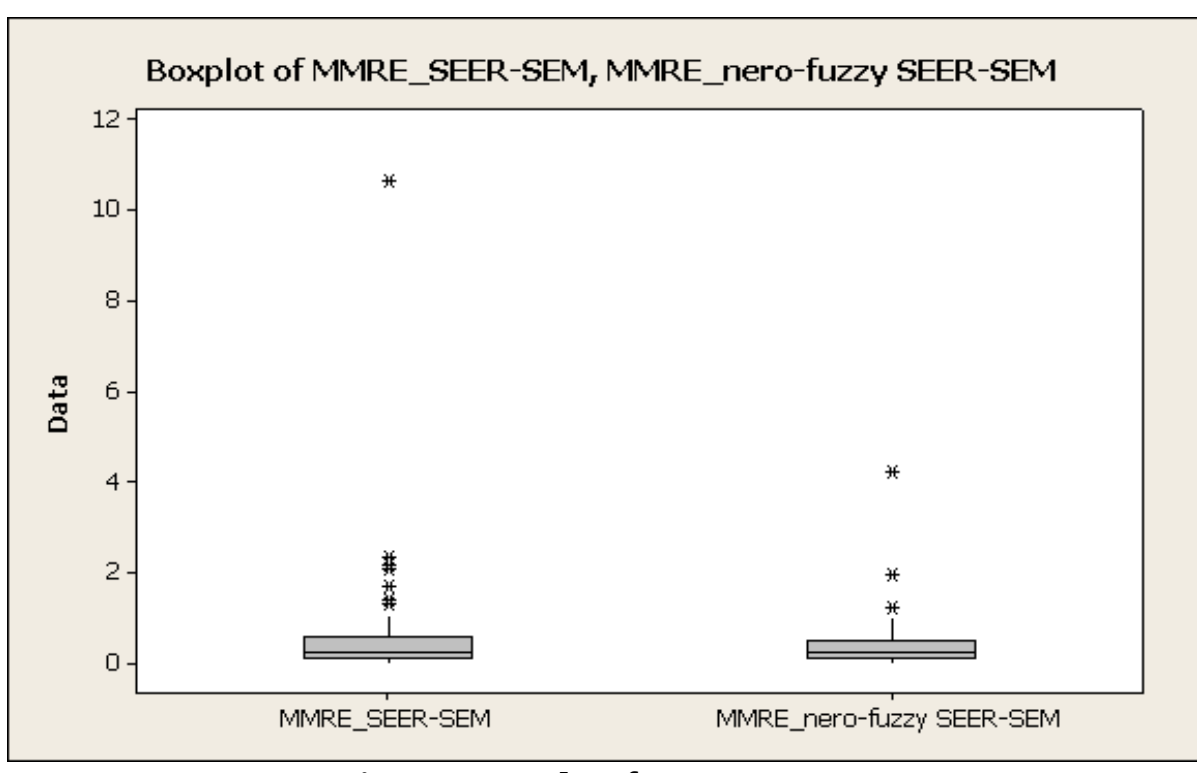

**Fig.5.** Boxplot for MMRE

**Table 1.** shows that the proposed neuro-fuzzy SEER-SEM model improves the original SEER-SEM model by 18% based on the MMRE criterion. Moreover, the values of MdMRE, PRED(30) and PRED(50) have been improved by 2%, 3% and 5%, respectively. Furthermore, we see significant improvement in the original SEER-SEM based on the MSE criterion. To better evaluate the significance of the proposed neuro-fuzzy SEER-SEM model, the Mann-Whitney U test was used. The p value reported is 0.0183. This indicates that the proposed model is significant at the 95% confidence level.

**Fig. 4.** and **Fig. 5.** also confirm the significance of the proposed model. **Fig. 4.** shows the interval plot of the MMRE for both models. The centre of the interval represents the MMRE value. Upper and lower edges represent the maximum and minimum values at 95% confidence interval. Regarding interval plots, the shorter the width of the interval is, the better the model. This shows that the prediction accuracy of the proposed neuro-fuzzy SEER-SEM is better than the original SEER-SEM model. In **Fig. 5.**, we see that in the SEER-SEM model, there are more points outside the Boxplot upper bound.  This indicates that the neuro-fuzzy model is better.

## 4. Threats to Validity

One of the main threats that might have affected the validity of this work is the scarce of the projects with SEER-SEM parameters. This is because SEER-SEM is a proprietary tool and SEER-SEM projects are not available online. For this reason, COCOMO projects were transformed to SEER-SEM and this indeed deteriorated the quality of the projects. Another threat we have encountered was the limited number of the projects used in this investigation. The accuracy of the model would have increased if the number of the projects was greater.

The large number of inputs to the model also has an adverse impact on the accuracy of the results. Limiting the number of the model's inputs not only decrease the complexity of the model, but also increases the accuracy.

## 5. CONCLUSIONS

Software engineering practitioners have always pursued the accuracy of software effort estimation for reducing costs, avoiding management risks, and achieving timely delivery. Through the continuous endeavor of researchers, various models and methodologies have been developed and introduced in software effort estimation. The main techniques adopted for effort estimation are briefly introduced in this article; these models are classified as experience-based, learning-oriented, model-based, regression-based, and composite techniques. Although many methodologies have been developed and adopted by practitioners, several significant difficulties still exist during software effort estimation, including the non-linear relationship between software size and estimation factors as well as the fact that software processes and techniques are evolving rapidly.

One of the techniques used by software effort estimation is soft computing, which assists in improving the estimation performance with its attractive and unique features. Specifically, fuzzy logic and neural networks are capable of effectively dealing with imprecise and uncertain information in addition to the complex, non-linear relationships of parameters. However, there are also shortcomings to the use of fuzzy logic and neural networks. For instance, a fuzzy system with a significant amount of complex rules cannot necessarily guarantee that the results will be meaningful, and the if-then rules are not adequately flexible for dealing with external changes. Moreover, neural networks contain the inherent feature of operating like a “black box”, which makes it difficult to prove that the model is working to the expectations of users. Thus, the neuro-fuzzy approach contains the advantages of fuzzy logic and neural networks as well as limits the disadvantages of these two techniques.

The proposed framework in this study is a combination of the machine-learning technique and the algorithmic effort estimation model, SEER-SEM. This framework is based on the unique architecture of the neuro-fuzzy model; in particular, ANFIS is a neuro-fuzzy technique adopted by the model. The neuro-fuzzy features of the model provide it with the advantages of strong adaptability with the capability of learning, less sensitivity for imprecise and uncertain inputs, and strong knowledge integration. On the whole, these techniques provide a good generalization for the proposed estimation model.

The aims of this research are to evaluate the prediction performance of the proposed neuro-fuzzy model with SEER-SEM in software estimation practices and to apply the proposed architecture that combines the neuro-fuzzy technique with different algorithmic models. Overall, the evaluation results indicate that estimation with our proposed neuro-fuzzy model containing SEER-SEM is more efficient than the estimation results that only use the SEER-SEM algorithm.

In this work, four different evaluation criteria have been used. These include the MMRE, MdMRE, PRED and MSE. Results show that the proposed model outperforms the original SEER-SEM model in the four criteria. The non-parametric Mann-Whitney U test was also used and results show that the proposed model is statistically significant at 95% confidence level.

Although several studies have already attempted to improve the general soft computing framework, there is still room for future work. First, the algorithm of the SEER-SEM effort estimation model is more complex than that of the COCOMO model. Prior research that combines neuro-fuzzy techniques with the COCOMO model demonstrates greater improvements in the prediction performance. Hence, the proposed general soft computing framework should be evaluated with other complex algorithms. Secondly, the datasets in our research are not from the original projects whose estimations are performed by SEER-SEM. When the SEER-SEM estimation datasets are available, more cases can be completed effectively for evaluating the performance of the neuro-fuzzy model. Also, future work will include studying the importance of each of the model’s inputs to see how much it is statistically significant.

In summary, this research demonstrates that combining the neuro-fuzzy model with the SEER-SEM effort estimation algorithm produces unique characteristics and performance improvements. Effort estimation using this framework is a good reference for the other popular estimation algorithmic models.

## APPENDIX A

| Parameters | SEER-SEM Rating | COCOMO Rating | Drivers/Factors |
|---|---|---|---|
| ACAP | *VLo-* | | ACAP |
| | VLo | VLo | |
| | Low | Low | |
| | Nom | Nom | |
| | Hi | Hi | |
| | VHi | VHi | |
| AEXP | VLo | VLo | APEX |
| | | *Low* | |
| | Low | Nom | |
| | Nom | Hi | |
| | Hi | VHi | |
| | *VHi* | | |
| PCAP | *VLo-* | | PCAP |
| | VLo | VLo | |
| | Low | Low | |
| | Nom | Nom | |
| | Hi | Hi | |
| | VHi | VHi | |
| LEXP | VLo | VLo | LTEX |
| | Low | Low | |
| | Nom | Nom | |
| | *Hi* | | |
| | VHi | Hi | |
| | XHi | VHi | |
| DEXP | VLo | VLo | PLEX |
| | Low | Low | |
| | Nom | Nom | |

| Parameters | SEER-SEM Rating | COCOMO Rating | Drivers/Factors |
|---|---|---|---|
| | ***Hi*** | | |
| | **VHi** | **Hi** | |
| | **XHi** | **VHi** | |
| **TEXP** | **VLo** | **VLo** | **PLEX** |
| | **Low** | **Low** | |
| | **Nom** | **Nom** | |
| | ***Hi*** | | |
| | **VHi** | **Hi** | |
| | **XHi** | **VHi** | |
| **MODP** | ***VLo*** | | **PMAT** |
| | **Low** | **VLo** | |
| | **Nom** | **Low** | |
| | **Hi** | **Nom** | |
| | **VHi** | **Hi, VHi, XHi** | |
| **TOOL** | **VLo** | **VLo** | **TOOL** |
| | **Low** | **Lo** | |
| | **Nom** | **Nom** | |
| | ***Nom+*** | | |
| | **Hi** | **Hi** | |
| | ***Hi+*** | | |
| | **VHi** | **VHi** | |
| **MULT** | **Nom** | **VHi, XHi** | **SITE** |
| | **Hi** | **Nom, Hi** | |
| | **VHi** | **Low** | |
| | **XHi** | **VLo** | |
| **DSVL TSVL** | ***Low*** | | **PVOL** |
| | **Nom** | **Low** | |
| | **Hi** | **Nom** | |
| | **VHi** | **Hi** | |
| | **XHi** | **VHi** | |
| **SPEC** | **VLo** | **VLo** | **RELY** |
| | **Low** | **Low** | |
| | **Nom** | **Nom** | |
| | **Hi** | **Hi** | |
| | **VHi** | **VHi** | |
| **REUS** | | ***Low*** | **RUSE** |
| | **Nom** | **Nom** | |
| | **Hi** | **Hi** | |
| | **VHi** | **VHi** | |
| | **XHi)** | **XHi** | |
| **APPL** | | **VLo** | **CPLX** |
| | **Low** | **Low** | |
| | | *Nom* | |
| | **Nom** | **Hi** | |
| | **Hi** | **VHi** | |

| Parameters | SEER-SEM Rating | COCOMO Rating | Drivers/Factors |
|---|---|---|---|
| | | ***XHi*** | |
| **MEMC** | **Nom** | **Nom** | **STOR** |
| | **Hi** | **Hi** | |
| | **VHi** | **VHi** | |
| | **XHi** | **XHi** | |
| **TIMC** | **Nom** | **Nom, Hi** | **TIME** |
| | **Hi** | **VHi** | |
| | **VHi** | **XHi** | |
| | ***XHi*** | | |
| **Staffing** | **VLo** | **VLo** | **CPLX** |
| | **Low** | **Low** | |
| | **Nom** | **Nom** | |
| | ***Nom+*** | | |
| | **Hi** | **Hi** | |
| | **VHi** | **VHi** | |
| | ***VHi+*** | | |
| | **XHi** | **XHi** | |
| **TURN** | **VLo** | **Low** | **TURN (COCOMO 81 cost driver)** |
| | **Low, Nom** | **Nom** | |
| | **Hi, VHi** | **Hi** | |
| | | ***VHi*** | |
| **DSVL** | **Low** | **Low** | **VMVH (COCOMO 87 Cost Driver)** |
| | **Nom** | **Nom** | |
| | **Hi** | **Hi** | |
| | **VHi** | **VHi** | |
| | **EHi** | | |
| **TSVL** | **Low** | **Low** | **VMVT (COCOMO 87 Cost Driver)** |
| | **Nom** | **Nom** | |
| | **Hi** | **Hi** | |
| | **VHi** | **VHi** | |
| | **EHi** | | |